\documentclass[12pt,a4paper,english,superscriptaddress,aps]{revtex4}
\usepackage{amssymb}
\usepackage{amsmath} 
\usepackage{slashed}
\usepackage{graphicx}
\usepackage{babel}
\usepackage{hyperref}
\usepackage{color}
\usepackage{comment}
\makeatletter\AtBeginDocument{\let\@elt\relax}\makeatother

\begin{document}

\title{Lorentz violating quadratic gravity}

\author{R. B. Alfaia}
\email[]{ramon.alfaia@icen.ufpa.br}
\affiliation{Faculdade de F\'{i}sica, Universidade Federal do Par\'{a}, 66075-110, Bel\'{e}m, Par\'a, Brazil}

\author{Willian Carvalho}
\email[]{willian.carvalho@icen.ufpa.br}
\affiliation{Faculdade de F\'{i}sica, Universidade Federal do Par\'{a}, 66075-110, Bel\'{e}m, Par\'a, Brazil}

\author{A. C. Lehum}
\email[]{lehum@ufpa.br}
\affiliation{Faculdade de F\'{i}sica, Universidade Federal do Par\'{a}, 66075-110, Bel\'{e}m, Par\'a, Brazil}
	
\author{J. R. Nascimento} 
\email[]{jroberto@fisica.ufpb.br}
\affiliation{Departamento de F\'{\i}sica, Universidade Federal da Para\'{\i}ba\\
Caixa Postal 5008, 58051-970, Jo\~ao Pessoa, Para\'{\i}ba, Brazil}

\author{A. Yu. Petrov}
\email[]{petrov@fisica.ufpb.br}
\affiliation{Departamento de F\'{\i}sica, Universidade Federal da Para\'{\i}ba\\
Caixa Postal 5008, 58051-970, Jo\~ao Pessoa, Para\'{\i}ba, Brazil}

\author{P. J. Porf\'{i}rio}\email[]{pporfirio@fisica.ufpb.br} \affiliation{Departamento de F\'{\i}sica, Universidade Federal da Para\'{\i}ba\\ 	Caixa Postal 5008, 58051-970, Jo\~ao Pessoa, Para\'{\i}ba, Brazil}

\begin{abstract}
In this paper we explore the perturbative renormalization and  study the classical dynamics of the bumblebee model coupled to quadratic gravity, a theoretical setting that allows the violation of Lorentz symmetry. Such a violation arises from a vector field whose potential is engineered to induce a nonzero vacuum expectation value (VEV), thereby leading to the emergence of a preferred direction in spacetime and, consequently, to the spontaneous breaking of Lorentz symmetry.
Working in dimensional regularization and expanding the metric around flat space, we compute the one-loop divergent parts of the two-point functions of the bumblebee and graviton fields, with special emphasis on the role of Lorentz-violating insertions in internal lines. These results determine the counterterms required to renormalize the gravitational and bumblebee sectors in the presence of a preferred background direction, and make explicit how Lorentz-violating interactions feed back into the UV structure of quadratic gravity. On the classical side, we derive the field equations and identify exact solutions supported by bumblebee backgrounds. In particular, we show that the Schwarzschild geometry remains an exact solution for an appropriate bumblebee configuration, even in the presence of non-minimal couplings. We close with a discussion of the operator content suggested by the one-loop structure and of prospective extensions to cosmological and less symmetric backgrounds.

\end{abstract}

\maketitle

\section{Introduction}

The idea that gravity could be consistently formulated as a quantum field theory has been explored from several complementary perspectives. One line of investigation has focused on constructing renormalizable extensions of general relativity by introducing curvature-squared terms into the action~\cite{Stelle:1976gc,Tomboulis:1977jk,Odintsov:1991nd,Salvio:2014soa,Einhorn:2014gfa,Salvio:2017qkx,Narain:2013eea}. Such higher-derivative formulations are known to improve the UV behavior of the theory and have been repeatedly proposed as potential frameworks for quantum gravity (see also  reviews~\cite{Buchbinder:2017lnd,Donoghue:2021cza}). Nevertheless, Einstein’s original theory without these modifications remains nonrenormalizable~\cite{tHooft:1974toh,Deser:1974zzd,Deser:1974cy}, and its predictive power is restricted to that one of an effective field theory, valid below the Planck scale~\cite{Donoghue:1994dn,Burgess:2003jk,Donoghue:2017pgk}.

At the same time, higher-derivative theories are not free from conceptual difficulties. Their particle spectra often include non-physical states, such as ghosts or tachyons, which threaten unitarity or stability. A variety of studies have addressed these issues from different angles~\cite{Holdom:2015kbf,Salvio:2018crh,Anselmi:2018ibi,Alvarez-Gaume:2015rwa,Donoghue:2021cza,Buccio:2024hys}, suggesting that while ghostlike modes might be tamed within certain formulations~\cite{Salvio:2014soa}, tachyonic instabilities usually signal more severe inconsistencies. Even so, as emphasized by Donoghue and Menezes~\cite{Donoghue:2021cza}, there is no definitive consensus regarding the presence or absence of such pathologies, leaving the fundamental status of quadratic gravity an open question that requires further theoretical clarification.

A particularly intriguing scenario emerges when quadratic curvature invariants are treated as the sole building blocks of the gravitational sector~\cite{Aoki:2021skm,Alvarez-Luna:2022hka,Aoki:2024jhr}. In this context, the absence of a fundamental mass scale leads to a renormalizable theory characterized by a graviton kinetic term with four derivatives and a propagator scaling as $1/p^4$~\cite{Buoninfante:2023ryt}, the so-called agravity. A notable outcome of this framework is the potential dynamical generation of the Planck scale at the quantum level~\cite{Gialamas:2020snr,Salvio:2014soa}. This occurs through a non-minimal coupling between a scalar field and gravity, $\xi \phi^2 R$, which induces a nonzero vacuum expectation value $\langle \phi \rangle \neq 0$ via radiative corrections, thereby setting the gravitational scale.

A compelling line of research in modified theories of gravity concerns the extension of general relativity to scenarios in which Lorentz symmetry is not preserved. As first emphasized in Ref.~\cite{KosSam}, the breaking of Lorentz invariance may naturally emerge from a spontaneous symmetry breaking (SSB) mechanism operating in the low-energy limit of a more fundamental framework, the string theory being a prominent example. Within this context, nontrivial vacuum configurations of vector or tensor fields can arise as the minima of a potential, thereby providing a dynamical origin for Lorentz-violating (LV) background fields. The spontaneous realization of Lorentz symmetry violation (LSV) in curved spacetimes is particularly attractive: not only does it furnish a consistent theoretical mechanism, but it also removes the usual constraint that LV vectors or tensors must be constant, a condition often imposed in flat spacetime treatments, but creating fundamental difficulties in curved space-times (see the discussion in \cite{KosLi} and references therein).

One of the most widely studied realizations of this idea is the so-called bumblebee model~\cite{KosGra,KosGra2,KosLi}, in which a vector field acquires a nonzero vacuum expectation value due to the presence of a potential term, while its dynamics are governed by a Maxwell-type kinetic structure. When coupled to gravity — possibly together with other matter fields — the resulting theory is known as bumblebee gravity. This framework has been extensively investigated in a variety of settings, including the analysis of modified black hole solutions~\cite{Bertolami,Casana2017,ourBH1,ourBH2}, cosmological dynamics~\cite{Capelo,Maluf2021}, G\"odel and G\"odel-type universes~\cite{ourgodel,gtype}, and even wormhole geometries~\cite{Ovgun}. Linearized versions of the theory have also been used to study modified dispersion relations~\cite{Maluf2014}. Beyond classical solutions, perturbative aspects of bumblebee gravity have also been explored, with explicit computations of quantum corrections performed in the metric–affine formalism~\cite{ourbumb1,ourbumb2,Lehum:2024ovo,AraujoFilho:2025hkm}. These results provide further insights into the structure and consistency of LV gravitational dynamics.

Given that the bumblebee field provides a natural framework for introducing LSV into gravitational models, it is a natural progression to extend the scale-invariant constructions discussed in Refs.~\cite{Salvio:2014soa,Salvio:2017qkx} by incorporating operators built from the bumblebee sector. 
A particularly noteworthy development in this direction was presented in Ref.~\cite{Lehum:2024wmk}, where it was demonstrated that coupling the bumblebee field to agravity can trigger the Coleman--Weinberg (CW) mechanism~\cite{Coleman:1973jx}, leading to dynamical Lorentz symmetry breaking within a scale-free framework.

It is important to emphasize that the present work is a different extension of the bumblebee-gravity framework. In Ref.~\cite{Lehum:2024wmk}, the gravitational sector was restricted to the scale-free agravity regime, involving only marginal curvature-squared operators, and Lorentz symmetry breaking was generated dynamically through radiative corrections via the Coleman--Weinberg mechanism. In contrast, the theory considered here also contains relevant gravitational operators, in particular the Einstein--Hilbert term, in addition to the marginal quadratic-curvature terms. As a consequence, the graviton propagator displays a richer pole structure, including the massless spin-2 pole together with the massive spin-2 and spin-0 poles characteristic of quadratic gravity. Moreover, in the bumblebee sector Lorentz symmetry is already spontaneously broken at tree level by the choice of a potential whose minimum occurs at \(\mathcal{B}_\mu \mathcal{B}^\mu=\pm b^2\). Therefore, while the two models share the same general motivation of coupling a bumblebee field to a renormalizable gravitational framework, the present analysis addresses a different perturbative regime, with a different propagator structure, a different pattern of Lorentz-symmetry breaking, and a different set of counterterms required for the renormalization of the effective theory.

One further motivation for studying this setup is the existence of nontrivial \emph{exact} solutions in bumblebee gravity. It is therefore natural to ask which geometries continue to be valid once the bumblebee sector is embedded into quadratic gravity and, in particular, how the solution space is affected by the non-minimal couplings introduced in Sec.~\ref{sec11}. This extends earlier investigations of exact solutions in related LV gravitational models~\cite{Bertolami,Casana2017,ourgodel,Delhom:2019wcm,ourbumb1,ourbumb2} to the present renormalizable framework.

Guided by the motivations discussed above, in this paper we carry out a twofold analysis of the bumblebee extension of quadratic gravity. First, we establish the perturbative framework by expanding around flat spacetime and determining the propagators and interaction vertices in the presence of the leading non-minimal couplings. We then compute the one-loop divergent parts of the bumblebee and graviton two-point functions, identifying the LV structures induced by gravitational fluctuations and extracting the counterterms required for renormalization when a preferred direction $b_\mu$ is present. Second, we address the classical sector by deriving the field equations and searching for exact solutions supported by bumblebee backgrounds; in particular, we show that the Schwarzschild geometry remains an exact solution for a suitable bumblebee configuration even at nonzero non-minimal couplings.

The structure of the paper looks as follows. In  Section 2, we introduce the Lagrangian of our theory and write down the propagators and the relevant vertices. In Section 3, we calculate the radiative corrections to two-point functions of the bumblebee and graviton fields, and in Section 4, we discuss exact solutions in our model. In Section 5, we summarize our results.

Throughout this paper, we use natural units $c=\hbar=1$.

\section{The bumblebee-quadratic gravity Lagrangian}\label{sec11}

The bumblebee model was proposed to investigate the possible spontaneous violation of Lorentz symmetry~\cite{KosGra}. It is defined by the action
\begin{equation}
S_B = \int d^4x \left\{ -\frac{1}{4}\mathcal{B}^{\mu\nu}\mathcal{B}_{\mu\nu} - V\left(\mathcal{B}^\mu \mathcal{B}_\mu \mp b^2 \right) \right\},
\end{equation}
where $\mathcal{B}_{\mu\nu} = \partial_{\mu} \mathcal{B}_{\nu} - \partial_{\nu} \mathcal{B}_{\mu}$, and the potential $V(\mathcal{B}^{\mu}\mathcal{B}_{\mu} \mp b^2)$ is chosen to induce a nonzero vacuum expectation value (VEV) for the bumblebee field. This leads to the emergence of a preferred direction in spacetime, resulting in spontaneous Lorentz symmetry breaking. It is worth mentioning that, first, the Maxwell-like kinetic term is necessary to avoid propagating ghost modes \cite{Chkareuli}, second, the bumblebee model is characterized by the presence of a potential displaying an infinite set of minima, being a consistent alternative to Einstein-aether gravity \cite{Jacobson} where the Lorentz symmetry breaking is introduced through the constraint instead of the potential term.

A common choice for the potential is $V = \lambda (\mathcal{B}^\mu \mathcal{B}_\mu \mp b^2)^2$, where the sign $\mp$ allows for both space-like and time-like configurations of $\mathcal{B}^\mu$, with $b^2 > 0$. The potential is responsible for Lorentz symmetry breaking, and its minimum occurs when $b^{\mu} b_{\mu} = \pm b^2$, implying that the VEV of the bumblebee field is given by $\langle \mathcal{B}_{\mu} \rangle = b_{\mu}$.

The coupling of the bumblebee field to the gravitational field, including quadratic curvature terms, is described by the action
\begin{eqnarray}\label{eq01}
	\mathcal{S} &=& \int{d^4x}\sqrt{-g}\Big\{\beta R^2-\alpha R^{\mu\nu}R_{\mu\nu}-\frac{\gamma}{\kappa^2}R 
	-\frac{1}{4} \mathcal{B}^{\mu\nu} \mathcal{B}_{\mu\nu}-\frac{1}{2g_l}(\nabla^\mu \mathcal{B}_\mu)^2\nonumber\\
	&& - \lambda (\mathcal{B}^\mu \mathcal{B}_\mu-b^2)^2 
    + \xi_1 \left(\mathcal{B}^\mu \mathcal{B}^\nu - \frac{g^{\mu\nu}}{4}\mathcal{B}^2 \right) R_{\mu\nu} + \xi_2 \mathcal{B}^2 R  +\mathcal{L}_{GF} + \mathcal{L}_{FP} + \mathcal{L}_{CT} \Big\},
\end{eqnarray}
where $R$ denotes the Ricci curvature scalar, $R_{\mu\nu}$ represents the Ricci tensor and $\nabla_\mu$ is the covariant derivative. The parameters $\alpha$, $\beta$ and $\gamma$ characterize the gravitational sector of the model. The constants $\xi_1$ and $\xi_2$ are the coupling coefficients associated with the traceless and trace components of the interaction between the bumblebee field and the gravitational field, respectively. Furthermore, $\mathcal{L}_{GF}$ and $\mathcal{L}_{FP}$ denote the gauge-fixing term and the corresponding Faddeev-Popov ghosts Lagrangian of the gravitational sector, respectively, while $\mathcal{L}_{CT}$ represents the Lagrangian of counterterms.

Let us consider a perturbative expansion of the metric tensor around the flat Minkowski background, given by
\begin{equation}
g_{\mu\nu} = \eta_{\mu\nu} + \kappa h_{\mu\nu},
\end{equation}
where $\eta_{\mu\nu}$ denotes the Minkowski metric and $h_{\mu\nu}$ represents the perturbation, assumed to be small. 

At the same time, the full bumblebee field is expanded around the LV vacuum as
\begin{equation}\label{B-expanded}
\mathcal B_\mu=b_\mu+B_\mu,
\qquad b_\mu b^\mu=\pm b^2 ,
\end{equation}
where \(\mathcal B_\mu\) denotes the full field, \(b_\mu\) is the vacuum expectation value, and \(B_\mu\) is the quantum fluctuation, with \(\langle B_\mu\rangle=0\). Thus, the fixed vector \(b_\mu\) is not introduced a posteriori, but defines the vacuum about which the perturbative expansion is performed. In the following, the covariant action is written in terms of \(\mathcal B_\mu\), while the weak-field expansion and the loop calculations are expressed in terms of the fluctuation \(B_\mu\).

By adopting this approach, one can formulate the tree-level Lagrangian before gauge fixing represented as
\begin{equation}\label{lag}
  \mathcal{L}=\mathcal{L}_h+\mathcal{L}_b +\mathcal{L}_{nm}+ \mathcal{L}_V,
\end{equation}
\noindent where $\mathcal{L}_h$ denotes the quadratic kinetic term of the gravitational Lagrangian
\begin{eqnarray} \label{gravi_new}
    \mathcal{L}_h &=&  - \gamma \Big[h\,\partial_{\mu}\partial_{\nu} h^{\mu\nu}+2\partial_{\nu}h^{\mu\nu}\partial_{\rho}h_{\mu}^{\,\,\,\rho}+\frac{1}{4}h\Box h -\partial_{\mu}h_{\nu}^{\,\,\,\rho} \partial_{\rho}h^{\mu\nu}
    +\partial_{\rho}h \partial^{\nu}h_{\nu}^{\,\,\,\rho} + \frac{1}{2}\partial_{\rho}\partial_{\nu}h\,h^{\nu\rho} 
    \Big.
    \nonumber\\
     &&\Big.+\frac{1}{4}h^{\mu\nu} \Box h_{\mu\nu}- \frac{1}{2} \partial^{\rho} h_{\mu\rho}\partial_{\nu} h^{\mu\nu}
    \Big] + \alpha\,\kappa^{2}\Big[\Big(\partial^{\rho}\partial_{\mu}h^{\mu\nu}\partial_{\nu}\partial_{\rho}h+\partial^{\rho}\partial_{\mu}h^{\mu\nu}\Box h_{\nu\rho}-\frac{1}{2}\Big(\partial_{\mu}\partial_{\nu}h^{\mu\nu}\Big)^{2}
    \Big.
     \nonumber\\
     &&\Big.-\frac{1}{2}\partial^{\rho}\partial_{\mu}h^{\mu\nu}\partial_{\rho}\partial_{\sigma}h^{\,\,\,\sigma}_{\nu}\Big)-\frac{1}{4}(\Box h)^{2} -\frac{1}{4}\Box h^{\mu\nu}\Box h_{\mu\nu} +\beta\Big(\Box h-\partial^{\mu}\partial_{\rho}h^{\,\,\,\rho}_{\mu}\Big)^{2}\Big] \nonumber\\
     &&{-\frac{\kappa^2}{2\zeta_g}\partial_\mu h^{\rho\mu}\Box \partial^\nu h_{\rho\nu}}+ \mathcal{O}(h^{3}),
\end{eqnarray}
$\mathcal{L}_{nm}$ comprises terms involving the couplings $\xi_1$ and $\xi_2$,
\begin{eqnarray}
	\mathcal{L}_{nm} &=& \xi_1 \kappa \mathcal{B}^{\alpha}\mathcal{B}^{\beta}\left( 
	\partial^{\gamma}\partial_{\beta}h_{\alpha\gamma}-\partial_{\alpha}\partial_{\beta}h -\frac{1}{2}\Box h_{\alpha\beta}\right) 
	+ \kappa \left(\xi_2-\frac{\xi_1}{4}\right) \mathcal{B}^2 \left(
	\partial^{\gamma}\partial^{\beta}h_{\gamma\beta}-\Box h		
	\right)\nonumber\\
	&&+\xi_1 \kappa^2 \mathcal{B}^\alpha \mathcal{B}^\beta \Big(\frac{1}{2}\partial^\nu {h_{\alpha}}^\mu \partial_\nu h_{\beta\mu} 
	-\frac{1}{2}\partial^\nu h_{\alpha\mu} \partial^\mu h_{\beta\nu}
	+\frac{1}{4}\partial^\alpha h^{\mu\nu} \partial_\beta h_{\mu\nu}
	+\frac{1}{2}\partial^\nu h_{\alpha\beta} \partial^\mu h_{\nu\mu}\nonumber\\
	&& - \partial_\beta {h_{\alpha}}^{\mu} \partial^\nu h_{\mu\nu}
	-\frac{1}{4}\partial^\nu h_{\alpha\beta} \partial_\nu h
	+\frac{1}{2}\partial_\beta h_{\alpha\mu} \partial^\mu h
	+\Box {h_{\beta}}^{\mu} h_{\alpha\mu}
	-\partial^\nu \partial^\mu h_{\beta\mu} h_{\alpha\nu}
	-\frac{1}{4}\Box h_{\alpha\beta} h\nonumber\\
	&& -\partial_\beta \partial^\nu {h_{\nu}}^{\mu} h_{\alpha\mu}
	+\partial^\mu \partial_\beta h h_{\alpha\mu} 
	+\frac{1}{2}\partial^\nu \partial_\beta h_{\alpha\nu}h
	-\frac{1}{4}\partial_\alpha \partial_\beta h h
	+\frac{1}{2}\partial^\nu \partial^\mu h_{\alpha\beta}h_{\mu\nu}\nonumber\\
	&&	-\partial^\nu \partial_\beta {h_{\alpha}}^{\mu}h_{\mu\nu}
	+\frac{1}{2}\partial_\beta \partial^\alpha h^{\mu\nu}h_{\mu\nu}
	\Big)
	+\kappa^2 \left(\xi_2-\frac{\xi_1}{4}\right) \mathcal{B}^{\alpha}\mathcal{B}^{\beta}\left(\Box h h_{\alpha\beta}-\partial^\mu \partial^\nu h_{\mu\nu} h_{\alpha\beta}\right) \nonumber\\
	&& 
	+\kappa^2 \left(\xi_2-\frac{\xi_1}{4}\right) \mathcal{B}^2 \Big( \frac{3}{4} \partial^\alpha h^{\beta\mu} \partial_\alpha h_{\beta\mu}
	-\frac{1}{2} \partial^\alpha h^{\beta\mu} \partial_\mu h_{\beta\alpha}
	+\partial^\alpha h \partial^\mu h_{\mu\alpha}
	- \partial^\alpha h^{\alpha\beta} \partial^\mu h_{\beta\mu}\nonumber\\
	&&
	-\frac{1}{2}  \Box h h
	 -\frac{1}{4}  \partial^\alpha h \partial_\alpha h 
	+\frac{1}{2}  \partial^\alpha \partial^\beta h_{\alpha\beta} h
	+\Box h^{\mu\nu} h_{\mu\nu} 
	-2 \partial^\alpha \partial_\beta h_{\alpha\mu} h^{\mu\beta}
	+\partial^\alpha \partial^\beta h h_{\alpha\beta}
	\Big)\nonumber\\
	&&+\,\mathcal{O}(h^3),
\end{eqnarray}
$\mathcal{L}_b$ encompasses contributions of the second order in the bumblebee field, generated by its kinetic term: 
\begin{eqnarray}\label{qp}
\mathcal{L}_b&=& -\frac{1}{4}\mathcal{B}^{\mu\nu}\mathcal{B}_{\mu\nu}\,-\frac{1}{2}\partial^{\mu}\mathcal{B}_{\mu}\partial^{\nu}\mathcal{B}_{\nu}\,+\frac{\kappa}{2}\mathcal{B}^{\mu\nu}\mathcal{B}^{\tau}_{\,\,\mu}\left(h_{\nu\tau}-\frac{1}{4}\eta_{\nu\tau}h \right)              \nonumber\\
&&-\frac{\kappa}{4} \left(\partial^{\mu}\mathcal{B}_\mu \partial^{\nu}\mathcal{B}_{\nu}h\,+\,2 \mathcal{B}^{\mu}\partial^{\nu}\mathcal{B}_{\nu}\partial_{\mu}h\,-\,4\mathcal{B}^{\mu}\partial^{\nu}\mathcal{B}_{\nu}\partial^{\alpha}h_{\mu\alpha}-4\partial^{\mu}\mathcal{B}_{\mu}\partial^{\nu}\mathcal{B}^{\alpha}h_{\nu\alpha}\right) \nonumber\\
&&-\frac{\kappa^2}{32}\left(h^2-2h^{\mu\nu}h_{\mu\nu}\right)\mathcal{B}^{\alpha\beta}\mathcal{B}_{\alpha\beta}\,-\frac{\kappa^2}{2}\mathcal{B}^{\mu}\mathcal{B}^{\nu}\left(\partial^{\alpha}h_{\mu\alpha}\partial^{\beta}h_{\nu\beta}-\partial^{\alpha}h_{\mu\alpha}\partial^{\nu}h\,+\frac{1}{4}\partial_{\mu}h\partial_{\nu}h\right)\nonumber\\
&&-\frac{\kappa^2}{2}\mathcal{B}^{\mu}\partial^{\beta}\mathcal{B}_{\beta}\left(2\,\partial_{\nu}h^{\alpha\nu}h_{\mu\alpha}-\partial^{\nu}hh_{\mu\nu}-\partial^{\nu}h_{\mu\nu}h\,+\frac{1}{2}\partial_{\mu}hh\,+2\,\partial_{\alpha}h_{\mu\nu}h^{\nu\alpha}-\partial_{\mu}h^{\alpha\nu}h_{\alpha\nu}\right)\nonumber\\
&&-\frac{\kappa^2}{2}\partial^{\alpha}\mathcal{B}_{\alpha}\left(2\,\partial^{\mu}\mathcal{B}_{\nu}h^{\nu\beta}h_{\mu\beta}-\partial^{\mu}\mathcal{B}_{\nu}h^{\mu\nu}h-\frac{1}{4}\partial^{\mu}\mathcal{B}_{\mu}h^{\nu\beta}h_{\nu\beta}\,+\frac{1}{8}\partial^{\mu}\mathcal{B}_{\mu}h^{2}\right)\nonumber\\
&&-\frac{\kappa^2}{2}\mathcal{B}^{\mu}\partial^{\alpha}\mathcal{B}^{\nu}\left(2\,\partial^{\beta}h_{\mu\beta}h_{\nu\alpha}\,-\partial_{\mu}hh_{\nu\alpha}\right)-\frac{\kappa^2}{2}\partial^{\mu}\mathcal{B}^{\nu}\partial^{\alpha}\mathcal{B}^{\beta}h_{\mu\nu}h_{\alpha\beta}\,+\mathcal{O}(h^3),
\end{eqnarray}
and $\mathcal{L}_V$ stands for the bumblebee potential and its coupling to gravity:
\begin{eqnarray}
     \mathcal{L}_V&=&-\lambda\Big(\mathcal{B}_\mu \mathcal{B}^\mu-b^{2}\Big)^{2}-\kappa \lambda\Bigg\{2\Big[\mathcal{B}^\mu \mathcal{B}^\nu b^2 h_{\mu\nu}-(\mathcal{B}_\mu \mathcal{B}^\mu(\mathcal{B}^\nu \mathcal{B}^\alpha-b^\nu b^\alpha)-b^2b^\nu b^\alpha]h_{\nu \alpha}\nonumber\\
     &&+\frac{1}{4}(\mathcal{B}_\mu \mathcal{B}^\mu-b^2)^2h\Big]\Bigg\}-\kappa^2\lambda\Bigg\{\frac{1}{8}\Big(\mathcal{B}_\mu \mathcal{B}^\mu-b^{2}\Big)^{2}+\mathcal{B}^\mu \mathcal{B}^\nu b^2 h_{\mu\nu}h-\nonumber\\ &-&\frac{1}{4}\Big(\mathcal{B}_\mu \mathcal{B}^\mu-b^{2}\Big)^{2}h_{\alpha\beta}h^{\alpha\beta}\nonumber\\
     &&-\Big[\mathcal{B}_\mu \mathcal{B}^\mu(\mathcal{B}^\nu \mathcal{B}^\alpha-b^\nu b^\alpha)+b^2 b^\nu b^\alpha\Big]h_{\nu\alpha}h+\left( \mathcal{B}^\mu \mathcal{B}^\nu - b^\mu b^\nu \right)
\left( \mathcal{B}^\alpha \mathcal{B}^\beta - b^\alpha b^\beta \right)
h_{\mu\nu} \, h_{\alpha\beta}\nonumber\\
     &&+2\Big(\mathcal{B}_\mu \mathcal{B}^\mu\big(\mathcal{B}^\nu \mathcal{B}^\alpha - b^\nu b^\alpha\big)
+ b_\mu b^\mu\, b^\nu b^\alpha \Big)h_\nu^\beta h_{\alpha\beta}\Bigg\}+\,\mathcal{O}(h^3).
\end{eqnarray}
{We omit the weak-field expansion of the Faddeev--Popov Lagrangian $\mathcal{L}_{\mathrm{FP}}$, since within the one-graviton exchange approximation adopted in this work the gravitational ghost fields do not contribute to the loop calculations.}

In a certain sense, our study can be regarded as an analogue of the one conducted in ~\cite{KosGra3}, where the weak-field limit was examined in the standard Einstein gravity coupled to the bumblebee field, at the tree level.

The quadratic part of our Lagrangian (\ref{lag}) corresponds to the following propagators:
\begin{eqnarray}
\Delta^{\mu\nu}(p) &=& -\frac{i}{p^2} \left[ T^{\mu\nu} + L^{\mu\nu} \right]; \nonumber\\
\Delta_{\mu\nu\rho\sigma}(p) &=& i\Bigg[\frac{2}{\gamma}\left(\frac{P^{(2)}_{\mu\nu\rho\sigma}-2P^{(0)}_{\mu\nu\rho\sigma}}{p^2}-\frac{P^{(2)}_{\mu\nu\rho\sigma}}{p^2-M^2_2}+\frac{2P^{(0)}_{\mu\nu\rho\sigma}}{p^2-M^2_0}\right)\nonumber \\&& +\frac{\zeta_g}{\kappa^2 p^4} \Big(3P^{(0)}_{\mu\nu\rho\sigma}-\sqrt{3}\left(P^{(0-sw)}_{\mu\nu\rho\sigma}+P^{(0-ws)}_{\mu\nu\rho\sigma}\right)+P^{(0w)}_{\mu\nu\rho\sigma}\Big)\Bigg],
\end{eqnarray}
where $\zeta_g$ is the gravitational gauge-fixing parameter, which we set to zero for simplicity.  The quantities $\Delta^{\mu\nu}(p)$ and $\Delta_{\mu\nu\rho\sigma}(p)$ denote the propagators of the bumblebee and graviton fields, respectively.  The mass parameters $M_0^2$ and $M_2^2$ are defined as
\begin{equation}
M_0^2 = \frac{\gamma}{2\kappa^2}(3\beta - \alpha),
\qquad
M_2^2 = \frac{\gamma}{\alpha\,\kappa^2}.
\end{equation}
Here, $M_0$ corresponds to the effective mass associated with the scalar (spin-0) excitation of the metric, while $M_2$ is the effective mass of the tensor (spin-2) excitation.  Both these constants arise from the higher-derivative terms in the quadratic action and characterize the propagating modes of the gravitational sector.

The projectors are defined as follows:
\begin{eqnarray}
P^{(2)}_{\mu\nu\rho\sigma} &=& \frac{1}{2} T_{\mu\rho} T_{\nu\sigma} + \frac{1}{2} T_{\mu\sigma} T_{\nu\rho} - \frac{1}{D-1} T_{\mu\nu} T_{\sigma\rho}; \nonumber\\
P^{(1)}_{\mu\nu\rho\sigma} &=& \frac{1}{2} \left( T_{\mu\rho} L_{\nu\sigma} + T_{\mu\sigma} L_{\nu\rho} + L_{\mu\rho} T_{\nu\sigma} + L_{\mu\sigma}T_{\nu\rho}\right); \nonumber\\
P^{(0)}_{\mu\nu\rho\sigma} &=& \frac{1}{D-1} T_{\mu\nu} T_{\sigma\rho}; \nonumber\\
P^{(0w)}_{\mu\nu\rho\sigma} &=& L_{\mu\nu} L_{\sigma\rho};\nonumber\\
P^{(0-sw)}_{\mu\nu\rho\sigma} &=& \frac{1}{\sqrt{3}} T_{\mu\nu}L_{\sigma\rho}\nonumber\\
P^{(0-ws)}_{\mu\nu\rho\sigma} &=& \frac{1}{\sqrt{3}} L_{\mu\nu}T_{\sigma\rho}
\end{eqnarray}
where $T_{\mu\nu}$ and $L_{\mu\nu}$ are given by
\begin{eqnarray}
T_{\mu\nu} &=& \eta_{\mu\nu} - \frac{p_\mu p_\nu}{p^2}; \nonumber\\
L_{\mu\nu} &=& \frac{p_\mu p_\nu}{p^2}.
\end{eqnarray}
These projectors will be used in our calculations.

Before proceeding to the loop calculation, let us clarify the field content around the LV vacuum. The full bumblebee field is expanded as in Eq.~\eqref{B-expanded}. Fluctuations tangent to the vacuum manifold are identified with the Nambu--Goldstone sector associated with the spontaneous breaking of local Lorentz symmetry~\cite{KosGra2,Maluf:2015hda}, whereas the fluctuation normal to the constraint surface corresponds to the potential, or massive, mode. In the perturbative calculation performed below, the LV structures proportional to \(b_\mu\) are treated as insertions around this vacuum configuration.

The gravitational sector is also richer than that of Einstein--Maxwell theory. Indeed, because the action contains quadratic-curvature operators, the graviton propagator decomposes into the usual spin projectors and contains, besides the massless spin-2 pole, the massive spin-2 and spin-0 poles characterized by \(M_2\) and \(M_0\), respectively. The usual caveats associated with the higher-derivative spin-2 pole are inherited from quadratic gravity and are not the focus of the present work. The nonminimal couplings \(\xi_1\) and \(\xi_2\) then generate Lorentz-violating mixing structures and local insertions in the quadratic action. Our purpose is therefore not to establish an equivalence with Einstein--Maxwell theory, but rather to determine how the bumblebee vacuum modifies the UV counterterm structure of quadratic gravity.

\section{Radiative Corrections in LV quadratic gravity}\label{sec3}

In this section, we present the UV renormalization of the quadratic part of the effective action of the model.  We begin by computing the radiative corrections to the quadratic sector of the effective action for the bumblebee field,  and then proceed with evaluating the corresponding corrections for the gravitational field.
 
\subsection{The bumblebee field self-energy}

We begin by examining the renormalization of the quadratic sector of the effective action for the bumblebee field. To proceed, we introduce the field redefinitions \( B_\mu \rightarrow Z_B^{1/2} B_\mu \) and \( b \rightarrow Z_b^{1/2} b \), where the \( Z \)’s denote the renormalization constants, each expanded perturbatively. Thus, the Lagrangian takes the form  
\begin{eqnarray}\label{LB1}
      \mathcal{L}_B &=& -\frac{Z_B}{4} \mathcal{B}_{\mu\nu} \mathcal{B}^{\mu\nu} 
      - \frac{Z_B}{2 g_l} (\partial_\mu \mathcal{B}^\mu)(\partial_\nu \mathcal{B}^\nu)
      - \lambda \left(Z_B\, \mathcal{B}_\mu \mathcal{B}^\mu - Z_b\, b^2 \right)^{2}.
\end{eqnarray}

The classical potential attains its minimum at \( B^\mu = b^\mu \). Therefore, we expand the Lagrangian around this vacuum configuration by performing the shift \( \mathcal{B}^\mu \rightarrow B^\mu + b^\mu \). After doing so, and defining \( Z_B = 1 + \delta_B \), \( \lambda Z_B^2 = \lambda + \delta_\lambda \), and \( 2 \lambda (Z_b - Z_B) Z_B = \delta_{m^2} \), the renormalized Lagrangian can be written as  
\begin{eqnarray}\label{LB2}
      \mathcal{L}_B &=& -\frac{1}{4} B_{\mu\nu} B^{\mu\nu}
      - \frac{1}{2 g_l} (\partial_\mu B^\mu)(\partial_\nu B^\nu)
      - \lambda \left(4 b^\mu b^\nu B_\mu B_\nu - 2 b^2 B^2 \right) \nonumber\\
      && - \frac{\delta_B}{4} B^{\mu\nu} B_{\mu\nu}
      - \frac{\delta_{g_l}}{2 g_l} (\partial_\mu B^\mu)(\partial_\nu B^\nu)
      - 4 \delta_\lambda\, b^\mu b^\nu B_\mu B_\nu
      + \delta_{m^2}\, b^2 B^2 + \cdots,
\end{eqnarray}
\noindent where the ellipsis ($\cdots$) denotes higher-order interaction terms.

\begin{figure}[ht!]
	\includegraphics[angle=0 ,width=10cm]{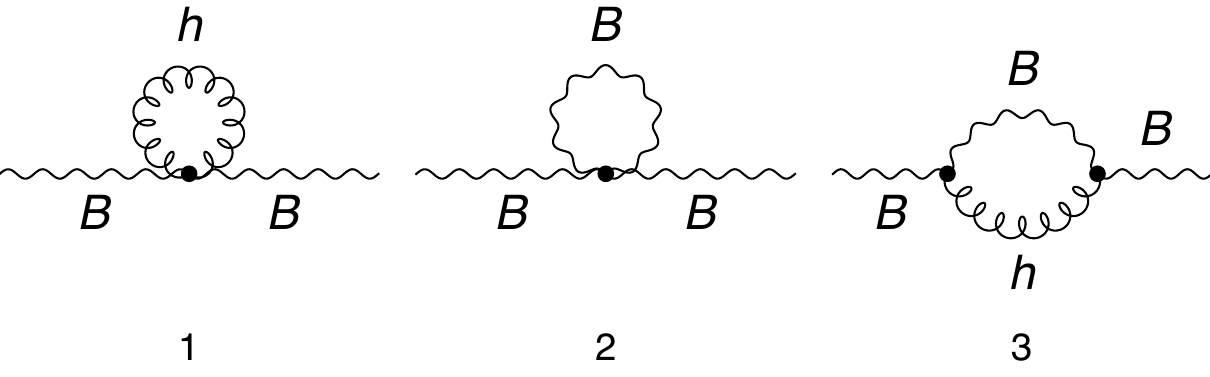}
	\caption{Bumblebee self-energy. Wavy and wiggly lines represent the bumblebee and graviton propagators, respectively.}
	\label{fig01}
\end{figure}

\begin{figure}[ht!]
	\includegraphics[angle=0 ,width=16cm]{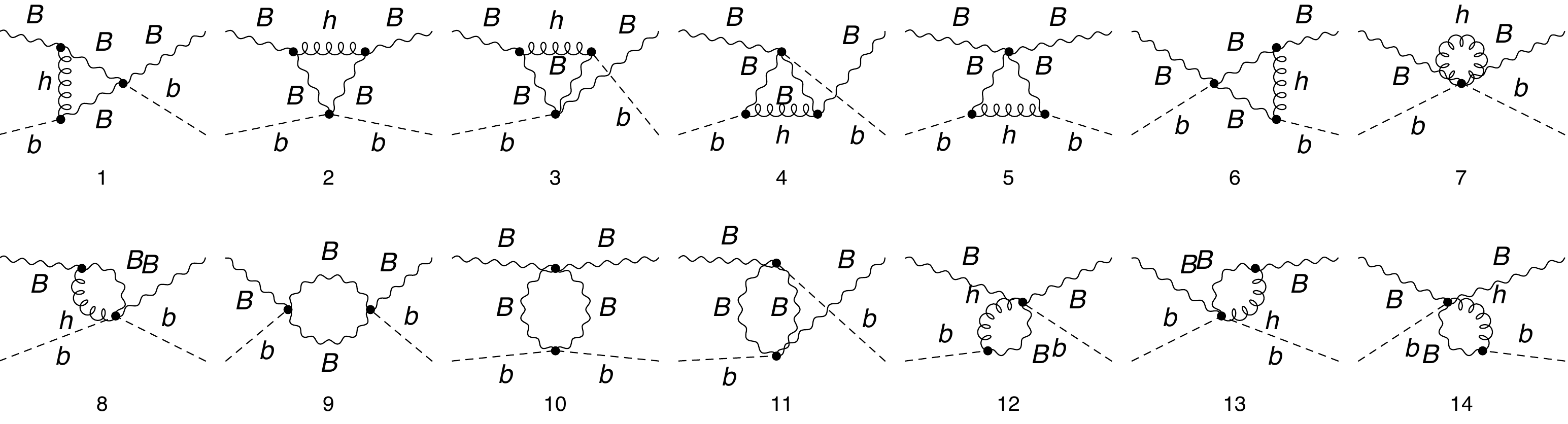}
	\caption{Bumblebee self-energy diagram with a Lorentz-violating vertex insertion, indicated by dashed lines.}
	\label{fig02}
\end{figure}

We compute the one-loop corrections to the self-energy of the bumblebee field, represented in Fig.~\ref{fig01}, as well as the corresponding LV contributions shown in Fig.~\ref{fig02}. By summing all these diagrams, we obtain the total one-loop contribution to the bumblebee field effective action, which can be expressed in the form
\begin{eqnarray}
\label{expansion}
S_B &=& \int d^4x \Big\{
-\frac{1}{4} \left( \delta_{T} -  \Gamma_T \right) B^{\mu\nu} B_{\mu\nu}
-\frac{1}{2} \left( \delta_{L} -  \Gamma_L \right) \partial_\mu B^\mu \partial_\nu B^\nu \\ \nonumber
&& + \left( \delta_{m^2} b^2- \Gamma_M \right)B^2 -4\left(\delta_\lambda-\Gamma_\lambda\right)b^\mu b^\nu B_\mu B_\nu \Big\},
\end{eqnarray}
where
\begin{eqnarray}
\Gamma_T &=& \frac{
\kappa^{2}
\left[
\left(
 \xi_{1}\!\left(
48\,\xi_{2}\,(2 g_l - 1)
+ 4\,\xi_{1}\,(5 g_l - 4)
- 5
\right) -132\,\xi_{2}
\right) M_{0}^{2}
+ 5\,\xi_{1}\,\big(\xi_{1}(2 g_l - 1) + 4\big)\,M_{2}^{2}
\right]
}{
288\pi^{2}\,\gamma\,\epsilon
}\nonumber;\\
\Gamma_L &=&\frac{
\kappa^{2}}{
96\pi^{2}\,\gamma\,g_l\,\epsilon
}\times\nonumber\\&\times&
\left[
\left(
36\,(1 - \xi_{2})
+ 4\,\xi_{1}\,(5 + \xi_{1})\,g_l
- 48\,(1 + 2\xi_{1})\,\xi_{2}\,g_l
- 8\,\xi_{1}\,(\xi_{1} - 6\xi_{2})\,g_l^{2}
- 3\,\xi_{1}
\right) M_{0}^{2}\right.\nonumber\\
&&\left.
+ 5\left(9 + \xi_{1}\big(6 + (2 + \xi_{1}(g_l - 2))\,g_l\big)\right) M_{2}^{2}
\right]
;\nonumber\\
\Gamma_M &=& \frac{
\kappa^{2}}{
256\pi^{2}\,\gamma\,\epsilon}
\Big[
\left(
16\,\xi_{2}\,(3\xi_{2}(3 + g_l) - 2)
+ 4\,\xi_{1}\,(6\xi_{2}(g_l - 1) - 1)
+ \xi_{1}^{2}\,(1 + 3g_l)
\right) M_{0}^{4} \nonumber\\
&& + 10\,(\xi_{1} - \xi_{1}^{2} + 2\xi_{2})\,M_{2}^{4} \Big]
-\frac{3\lambda^2 b^2}{\pi^2\epsilon}
-
\frac{
\lambda\kappa^{2} b^2}{
288\pi^{2}\,\gamma \,\epsilon}
\Big[10\,(32 - 21\xi_{1})\,\xi_{1}\,M_{2}^{2}\nonumber\\
&&+
\Big(
3\xi_{1}^{2}\,(29 + 3g_l^{2} - 4g_l)
+ 2\xi_{1}\,\big(12\xi_{2}(g_l - 1)(7 + 3g_l) + 38 - 9g_l\big)\nonumber\\
&&+ 72\xi_{2}\,\big(2\xi_{2}(55 + g_l(4 + g_l)) - 2 - g_l\big)
+ 9
\Big) M_{0}^{2}
\Big]
; \nonumber\\
\Gamma_\lambda &=& \frac{3\lambda^2}{\pi^2\, \epsilon}
+\frac{
\lambda \kappa^{2} }{
1152\,\epsilon\,\pi^{2}\,\gamma
}
\Big[40\,(22 - 15\xi_{1})\,\xi_{1}\,M_{2}^{2}+
\Big(
45
+ 182\,\xi_{1}
+ 201\,\xi_{1}^{2}
- 72\,\xi_{2}
\nonumber\\
&& - 408\,\xi_{1}\xi_{2}
+ 19728\,\xi_{2}^{2}
- 6\,(15 + \xi_{1} - 12\xi_{2})(\xi_{1} + 4\xi_{2})\,g_l
+ 45\,(\xi_{1} + 4\xi_{2})^{2}\,g_l^{2}
\Big) M_{0}^{2}
\Big].
\end{eqnarray} 

By imposing the finiteness requirement through the MS scheme, we find the counterterms to be $\delta_M=b^2 \Gamma_M$, $\delta_{T}=\Gamma_T$ and $\delta_{L}=\Gamma_L$.

A noteworthy structural outcome of the quadratic expansion, Eq.~(\ref{expansion}), is that the bumblebee sector is not closed under renormalization if one restricts the action to the Maxwell-type kinetic term alone. Indeed, the one-loop divergences contain a purely longitudinal structure proportional to $(\partial_\mu B^\mu)^2$, which cannot be absorbed by renormalizing the transverse Maxwell operator $-\tfrac14 B_{\mu\nu}B^{\mu\nu}$. Therefore, perturbative renormalizability requires the inclusion of the longitudinal counterterm
$\Delta \mathcal{L}_{\rm ct}=-\frac{\delta_L}{2}\,(\partial_\mu B^\mu)^2$  (or an equivalent operator basis), ensuring that the set of local operators is stable under radiative corrections. In the absence of this term (or in singular parameter limits where the longitudinal sector becomes degenerate), the UV divergences cannot be removed by a finite renormalization of the original couplings, signaling the loss of perturbative renormalizability.

Although a counterterm of the cosmological-term type, $\delta_\Lambda$, is physically admissible, in this work we restrict our analysis to counterterms proportional to $b^2$, neglecting higher-order contributions such as $b^4$.

\subsection{Graviton-bumblebee mixing at tree level}

\begin{figure}[ht!]
	\includegraphics[angle=0 ,width=5cm]{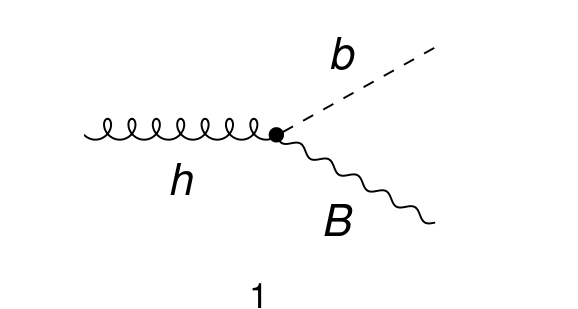}
	\caption{Graviton-bumblebee transmutation induced by the LV background.}
	\label{fig03}
\end{figure}

We proceed to compute the tree-level amplitude for the bumblebee–graviton interaction by evaluating the contributions from the LV vertices. The corresponding Feynman diagram is shown in Figure~\ref{fig03}. The resulting expression for $|\mathcal{M}|^2$ is given by
\begin{eqnarray}\label{hgo2B}
|\mathcal{M}|^2 &=&
\frac{\kappa^2}{64\,g_l^2}
\Bigg\{ 
\varepsilon^{(g)}_{\mu\nu}(p,\sigma)\, \Big[
8\left(\xi_1 g_l-1\right)
\, \big(p\cdot \epsilon(p,\rho)\big) b^\mu p^\nu
-4(\xi_1-4\xi_2)g_l\,
\big(b\cdot \epsilon(p,\rho)\big)\,
p^\mu p^\nu
\nonumber \\[6pt]
&&\quad
+\,4\,(b\cdot p)\left(
(1-2\xi_1 g_l)\,\big(p\cdot \epsilon(p,\rho)\big)\, 
\eta^{\mu\nu}
+2\xi_1 g_l\, 
\epsilon^\mu(p,\rho)\, p^\nu
\right)
\Big]
\nonumber \\[6pt]
&&\times
\varepsilon^{(g)\,*}_{\alpha\beta}(p,\sigma)\, \Big[
8\left(\xi_1 g_l-1\right)
\, \big(p\cdot \epsilon^*(p,\rho)\big)\,
 b^\alpha p^\beta
-4(\xi_1-4\xi_2)g_l\,
\big(b\cdot \epsilon^*(p,\rho)\big)\,
p^\alpha p^\beta
\nonumber \\[6pt]
&&\quad
+\,4\,(b\cdot p)\left(
(1-2\xi_1 g_l)\,\big(p\cdot \epsilon^*(p,\rho)\big)\, 
\eta^{\alpha\beta}
+2\xi_1 g_l\, 
\epsilon^{*\alpha}(p,\rho)\, p^\beta
\right)
\Big]
\Bigg\},
\end{eqnarray}
where $\varepsilon^{(g)}(p,\sigma)$ and $\epsilon(p,\rho)$ are the graviton polarization tensor and bumblebee polarization vector, respectively.

It is straightforward to verify that the squared amplitude in \eqref{hgo2B} vanishes identically for an external graviton on-shell. In fact, every term in $|\mathcal{M}|^2$ involves either a contraction of the graviton polarization tensor with the external momentum, $\varepsilon_{\mu\nu}(p)\,p^\nu$, or its trace, $\varepsilon^{\mu}{}_{\mu}(p)$, always multiplied by factors proportional to $p\cdot\varepsilon_{\pm}(p)$. Since physical graviton polarizations satisfy the transversality and tracelessness conditions, $p^\mu\varepsilon_{\mu\nu}(p)=0$ and $\varepsilon^{\mu}{}_{\mu}(p)=0$, while the bumblebee polarizations obey $p\cdot\varepsilon_{\pm}(p)=0$, each contribution vanishes separately. Therefore, the squared amplitude reduces to zero once the on-shell conditions are imposed.

In contrast, when this vertex is embedded in a loop, the amplitude (see Figure \ref{fig04}) does not vanish. The reason is that, in this case, the internal bumblebee lines are off-shell and must be accounted for through the full propagator rather than through the physical polarization conditions. Since the bumblebee field is treated as massless and all LV vertices are considered perturbative insertions, the completeness relation for the polarization sum introduces the axial-gauge type projector $\pi_{\mu\nu}(p,n)=\eta_{\mu\nu}-(p_\mu n_\nu+p_\nu n_\mu)/(p\cdot n)$. This projector does not annihilate momentum dependence, and contractions with the background vector $b^\mu$ generate nontrivial structures such as $(b\!\cdot p)^2$ and $b^2 p^2$ that survive after integration. Consequently, although $|\mathcal{M}|^2$ vanishes for external on-shell gravitons, loop diagrams involving internal bumblebee propagators give rise to nonzero contributions.

\subsection{The graviton self-energy}

\begin{figure}[ht!]
	\includegraphics[angle=0 ,width=8cm]{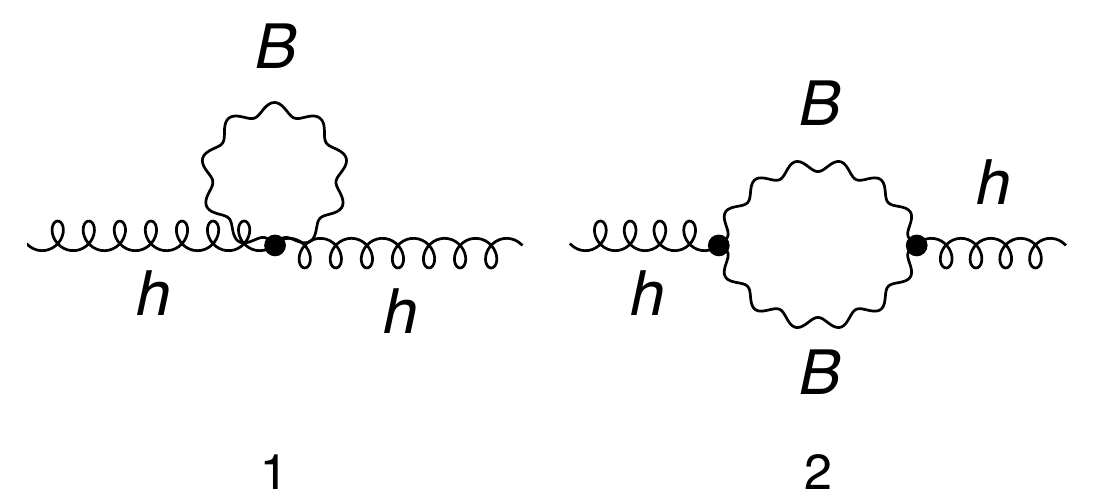}
	\caption{Bumblebee corrections to the graviton propagation.
    }
	\label{fig04}
\end{figure}

Our next step is to compute the quadratic part of the graviton effective action, that is, to evaluate the one-particle-irreducible (1PI) diagrams contributing to the graviton self-energy. The corresponding Feynman diagrams are shown in Fig.~\ref{fig04}. It is immediate to see that the contribution from diagram~\ref{fig04}.1 vanishes identically due to the massless nature of the $B$ field. The expression associated with diagram~\ref{fig04}.2 yields the contribution of the bumblebee fluctuations to the quadratic part of the graviton effective action, which can be written as
\begin{eqnarray}
    S_g^{(2)} &=& \int d^4x\sqrt{-g} \left[-\left(\delta_\alpha-\Gamma_\alpha\right) R^{\mu\nu}R_{\mu\nu}+\left(\delta_\beta-\Gamma_\beta\right)R^2-\frac{\delta_\gamma-\Gamma_\gamma }{\kappa^2}R\right],
\end{eqnarray}
where the superscript $^{(2)}$ denotes the quadratic part in the graviton field $h^{\mu\nu}$, and 
\begin{eqnarray}
\Gamma_\alpha &=& \frac{
7 - 10\,\xi_{1}\,(5 + g_l)
+ 5\,\xi_{1}^{2}\,(7 + 4g_l + g_l^{2})
}{
960\pi^{2} \,\epsilon};\nonumber\\
\Gamma_\beta &=&\frac{6
- 10\,\xi_{1}\,(5 + g_l)
- 40\,\xi_{2}\,(g_l - 3)
- 240\,\xi_{2}^{2}\,(3 + g_l^{2})
+ 5\,\xi_{1}^{2}\,(7 + 4g_l + g_l^{2})}{
3840\pi^{2}\,\epsilon};
\nonumber\\
\Gamma_\gamma &=&0.
\end{eqnarray}
Notice that the counterterms $\delta_\alpha$, $\delta_\beta$, and $\delta_\gamma$ are generated from Eq.~\eqref{gravi_new} through the field redefinition $h^{\mu\nu} \rightarrow \sqrt{Z_h}\, h^{\mu\nu}$,
and by identifying $\alpha+\delta_\alpha=Z_h\alpha$, $\beta+\delta_\beta=Z_h\beta$ and $\gamma+\delta_\gamma=Z_h\gamma$.

By imposing finiteness within the framework of the MS renormalization scheme, we obtain the counterterms
$\delta_\alpha = \Gamma_\alpha$, $\delta_\beta = \Gamma_\beta$, and $\delta_\gamma = 0$. It is worth emphasizing that, since the LV terms are treated as perturbations and the bumblebee field is massless at zeroth order in the LV parameters, there is no renormalization of the Einstein-Hilbert term.

\begin{figure}[ht!]
	\includegraphics[angle=0 ,width=16cm]{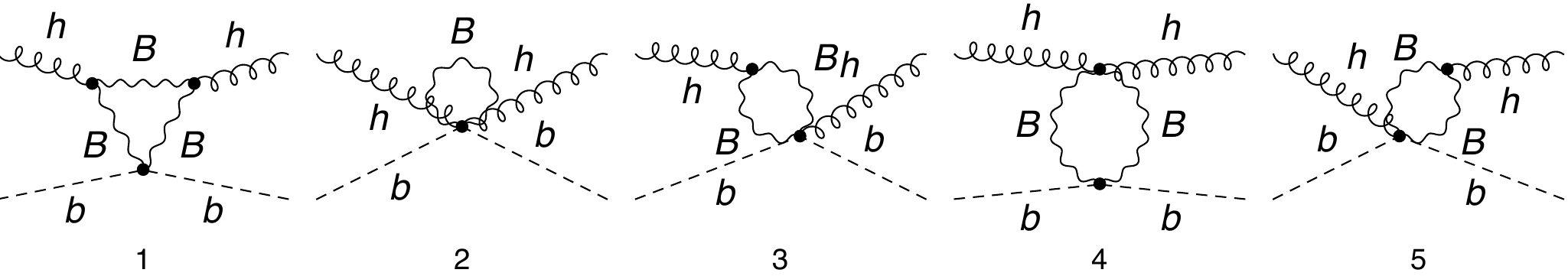}
	\caption{LV corrections to the graviton propagation.}
	\label{fig05}
\end{figure}

We now focus our attention on the LV corrections to the graviton self-energy. The relevant contributions are encoded in the Feynman diagrams displayed in Figure~\ref{fig05}. The LV correction to the quadratic part of the graviton effective action can be expressed as  
\begin{eqnarray}
    S_g^{(2)} &=& \int d^4x \sqrt{-g}\Bigg[\left(\delta_{\xi_1}-\frac{
\lambda\,
\big[ \xi_{1}\,(7 + 4g_l + g_l^{2})
- (5 + g_l) \big]}{24\pi^{2}\,\epsilon}\right) \left(b^\mu b^\nu - \frac{g^{\mu\nu}}{4}b^2 \right) R_{\mu\nu} \nonumber \\[6pt]
&&\quad+ \left(\delta_{\xi_2}-
\frac{ \lambda\, \big[ 12\,\xi_{2}\,(3 + g_l^{2}) + g_l - 3 \big]}{
96\pi^{2}\,\epsilon}
\right) b^2 R\Bigg],
\end{eqnarray}
where $\delta_{\xi_i}=Z_hZ_B\xi_i-\xi_i$, with $i=1,~2$. It is worth emphasizing that the insertions of LV vertices induce the renormalization of $\xi_1$, which is associated with the traceless part of the non-minimal coupling between the bumblebee and gravitational fields. In contrast, the trace part, proportional to $\xi_2$, corresponds to the LV correction to the Einstein-Hilbert term and appears only at the order $b^2$. We note also that the gravitational aether term $b^{\mu}b^{\nu}R_{\mu\nu}$ proposed in \cite{Carroll:2008pk} naturally arises within our calculations.

\section{Classical solutions}

Our bumblebee-gravity model (\ref{eq01}) can also be studied at the tree level. In this case, we disregard the gauge-fixing, bumblebee longitudinal term, ghosts and counterterms to obtain the equations of motion.

Let our Lagrangian looks like
\begin{eqnarray}
{\cal L}&=&\sqrt{-g}\Big(-\frac{1}{4}\mathcal{B}_{\mu\nu}\mathcal{B}^{\mu\nu}-\frac{\lambda}{4}(\mathcal{B}^{\mu}\mathcal{B}_{\mu}\mp b^2)^2+\alpha R+\beta R^2+\gamma R_{\mu\nu}R^{\mu\nu}+\nonumber\\&+&
\xi_1 \mathcal{B}^{\mu}\mathcal{B}^{\nu}R_{\mu\nu}+\xi_2 \mathcal{B}^2R
\Big),
\end{eqnarray}
that is, Eq. (\ref{eq01}) where the gauge-fixing and ghost terms are omitted as will as counterterms since all these contributions are relevant only within the perturbative framework. The constant parameters are relabeled as $\alpha$, $\beta$, $\gamma$ for convenience reasons.

The equation of motion for the bumblebee field has the simple form being a generalization of those ones given in \cite{ourgodel}:
\begin{eqnarray}
\label{maxw}
\nabla_{\mu}\mathcal{B}^{\mu\nu}-\lambda(\mathcal{B}^{\rho}\mathcal{B}_{\rho}\mp b^2)\mathcal{B}^{\nu}+2\xi_1\mathcal{B}_{\mu}R^{\mu\nu}+2\xi_2\mathcal{B}^{\nu}R=0.
\end{eqnarray}
The equations of motion for the gravitational field look like (they display some similarity with \cite{Casana2017}, where, however, terms proportional to $\beta$ and $\gamma$ were absent):
\begin{eqnarray}
\label{eqmot}
&&\alpha(R_{\mu\nu}-\frac{1}{2}Rg_{\mu\nu})+4\beta[RR_{\mu\nu}-\frac{1}{4}R^2g_{\mu\nu}+(g_{\mu\nu}\Box-\nabla_{\mu}\nabla_{\nu})R]+\nonumber\\
&+&\gamma[-\frac{1}{2}R_{\rho\sigma}R^{\rho\sigma}g_{\mu\nu}+2R_{\mu}^{\phantom{\mu}\rho}R_{\nu\rho}+2R^{\rho\sigma}R_{\mu\rho\nu\sigma}+\Box R_{\mu\nu}+\frac{1}{2}g_{\mu\nu}\Box R-\nabla_{\mu}\nabla_{\nu}R]+\nonumber\\
&+&\xi_1\Big[\frac{1}{2}g_{\mu\nu}\mathcal{B}^{\alpha}\mathcal{B}^{\beta}R_{\alpha\beta}-2\mathcal{B}_{(\mu}\mathcal{B}^{\rho}R_{\nu)\rho}-\nabla^{\rho}\nabla_{(\mu}(\mathcal{B}_{\nu)}\mathcal{B}_{\rho})+\frac{1}{2}g_{\mu\nu}\nabla^{\rho}\nabla^{\sigma}(\mathcal{B}_{\rho}\mathcal{B}_{\sigma})+\nonumber\\&+&
\frac{1}{2}\Box(\mathcal{B}_{\mu}\mathcal{B}_{\nu})-\nabla^{\rho}\nabla_{(\mu}\mathcal{B}_{\nu)}\mathcal{B}_{\rho}
\Big]+\nonumber\\
&+&\xi_2\Big[-\mathcal{B}^{\rho}\mathcal{B}_{\rho}G_{\mu\nu}+\nabla_{\mu}\nabla_{\nu}(\mathcal{B}^{\rho}\mathcal{B}_{\rho})-g_{\mu\nu}\Box(\mathcal{B}^{\rho}\mathcal{B}_{\rho})\Big]
=T_{\mu\nu}^B,
\end{eqnarray}
where $G_{\mu\nu}=R_{\mu\nu}-\frac{1}{2}Rg_{\mu\nu}$ is the standard Einstein tensor, and
\begin{eqnarray}
\label{tmn}
T_{\mu\nu}^B=\mathcal{B}_{\mu\alpha}\mathcal{B}_{\nu}^{\phantom{\nu}\alpha}-\frac{1}{4}g_{\mu\nu}\mathcal{B}_{\lambda\rho}\mathcal{B}^{\lambda\rho}-Vg_{\mu\nu}+2V^{\prime}\mathcal{B}_{\mu}\mathcal{B}_{\nu}\,,
\end{eqnarray}
is the standard (that is, $\xi$-independent) part of the energy-momentum tensor of the bumblebee field (cf. e.g. \cite{ourgodel}).

Let us solve these equations. We note that here, in the bumblebee sector, our aim consists in finding the minimum of the potential compatible with the equation of motion of the bumblebee field. As a first attempt, we can choose the purely radial $\mathcal{B}_{\mu}(r)=(0,\mathcal{B}_r(r),0,0)$, similar to \cite{Bertolami}. In this case, one easily arrives at $\mathcal{B}_{\mu\nu}=0$, moreover, it is shown in \cite{Bertolami}, that for the static spherically symmetric metric corresponding to the following line element:
\begin{eqnarray}
ds^2=e^{2\phi(r)}dt^2-e^{2\rho(r)}dr^2-r^2(d\theta^2+\sin^2\theta d\phi^2),
\end{eqnarray}
and for a certain form of $\mathcal{B}_r(r)$, that is, $\mathcal{B}_r(r)=be^{\rho(r)}$, where $b$ is constant, one has $\mathcal{B}^{\mu}\mathcal{B}_{\mu}=-b^2$ with $\mathcal{B}_{\mu}(r)$ being the vacuum. So, for this choice, $T_{\mu\nu}^B$ (\ref{tmn}) vanishes. Then, let us, for the sake of simplicity, consider the case $\xi_1=0$. In this case, our modified Einstein equation (\ref{eqmot}) reduces to
\begin{eqnarray}
\label{eqmot1}
&&\alpha(R_{\mu\nu}-\frac{1}{2}Rg_{\mu\nu})+4\beta[RR_{\mu\nu}-\frac{1}{4}R^2g_{\mu\nu}+(g_{\mu\nu}\Box-\nabla_{\mu}\nabla_{\nu})R]+\nonumber\\
&&+\gamma[-\frac{1}{2}R_{\rho\sigma}R^{\rho\sigma}g_{\mu\nu}+2R_{\mu}^{\phantom{\mu}\rho}R_{\nu\rho}+2R^{\rho\sigma}R_{\mu\rho\nu\sigma}+\Box R_{\mu\nu}+\frac{1}{2}g_{\mu\nu}\Box R-\nabla_{\mu}\nabla_{\nu}R]-\nonumber\\
&&-
\xi_2\mathcal{B}^{\rho}\mathcal{B}_{\rho}G_{\mu\nu}
=0.
\end{eqnarray}
This equation (where $\mathcal{B}_{\mu}$ is chosen to be in the radial form we discussed above) certainly admits various solutions. The simplest one is the Schwarzschild metric -- indeed, for this metric $R_{\mu\nu}=R=0$, and the bumblebee equation of motion (\ref{maxw}) is also satisfied. So, we succeeded in demonstrating that our modified Einstein equations, besides the trivial Minkowski solution, are solved at least by the Schwarzschild metric. Another interesting solution for the case $\xi_1$=0 is the static de Sitter metric for which one has $R_{\mu\nu\lambda\rho}=\frac{1}{a^2}(g_{\mu\lambda}g_{\nu\rho}-g_{\mu\rho}g_{\nu\lambda})$, where $a$ is a constant,  with $R_{\mu\nu}=\frac{3}{a^2}g_{\mu\nu}$ and $R=\frac{12}{a^2}$, in this case, at $\xi_1=0$, the equation (\ref{eqmot1}) is satisfied if the following algebraic relationship between constants of the theory holds: $-\frac{3\alpha}{a^2}+\frac{18\gamma}{a^4}-3\frac{\xi_2b^2}{a^2}=0$. In a certain sense, the nonminimal coupling, together with $\gamma$, in this case, is related to the cosmological constant $\Lambda=\frac{1}{a^2}$.

The natural question is whether one can have a nontrivial solution when both $\xi_1$ and $\xi_2$ are non-zero. There is a nontrivial solution in this case, that is, the {\it cylindrically} symmetric metric known as the Levi-Civita metric (see e.g. \cite{Racsko:2022lls} and references therein):
\begin{eqnarray}
ds^2=\rho^{4m}dt^2-\rho^{4m(m-1)}(d\rho^2+\rho^2d\phi^2)-dz^2.
\end{eqnarray}
This metric satisfies the vacuum Einstein equations $R_{\mu\nu}=0$. At the same time, if we choose the vacuum vector along the $z$ axis, $\mathcal{B}_{\mu}=(0,0,0,b)$ (so, $\mathcal{B}^{\mu}\mathcal{B}_{\mu}=-b^2$ as it must be in the space-like case), we see that, as for the Levi-Civita metric one easily finds $\Gamma^3_{\mu\nu}=\Gamma^{\mu}_{3\nu}=0$, the identity $\nabla_{\mu}\mathcal{B}_{\nu}=0$ is valid for our vacuum. Hence, all terms proportional to $\xi_1$ vanish as well, and we proved that this metric satisfies the modified Einstein equation for both $\xi_1,\xi_2\neq 0$.

Perhaps, some other metrics can be shown to solve these equations perturbatively, taking into account that solutions of Eqs. (\ref{maxw},\ref{eqmot}) at $\xi_{1,2}=0$ can be modified by small perturbations proportional to $\xi_{1,2}$. The cosmological issues related to this theory certainly deserve a separate study. Also, possible non-spherically-symmetric solutions (for example, Kerr metric or $pp$-waves) in our theory can be tested.

\section{Final Remarks}\label{summary}

Let us discuss our results. In this work, we have formulated a model displaying spontaneous Lorentz symmetry breaking within the framework of quadratic gravity. This model includes the terms of first and second orders in curvature, therefore being an extension of the model considered in \cite{Lehum:2024wmk}, by incorporating the most general renormalizable curvature sector.

Within the classical context, we derived the field equations and analyzed stationary, spherically symmetric configurations supported by bumblebee backgrounds. In particular, we showed that the Schwarzschild geometry remains an \emph{exact} solution of the full theory for an appropriate bumblebee configuration, even in the presence of non-minimal couplings.

Within the quantum context, we expanded the action to second order in fluctuations around flat spacetime and computed the one-loop two-point functions of the bumblebee and graviton fields. A notable structural outcome concerns the mixed graviton--bumblebee two-point function induced by a LV vertex insertion: for on-shell external legs, it vanishes, consistently with the transversality/tracelessness properties of the physical graviton polarization tensor. However, this kinematical decoupling does not prevent the corresponding mixed structures from contributing inside loop diagrams, since the internal lines are generically off-shell and the same cancellations no longer apply. At the same time, LV insertions generate, in the graviton sector, the aether-type operators advocated in Ref.~\cite{Carroll:2008pk}. Importantly, we find that these aether-like contributions arise with UV divergences, implying that they must be included as counterterms in a renormalized treatment and providing a concrete illustration of how spontaneous Lorentz violation feeds back into the UV operator content of quadratic gravity.

A further continuation of this study could consist, first, in testing more involved gravitational solutions within our model, second,
in studying the homogeneous and isotropic cosmological backgrounds and the resulting modifications of the Friedmann dynamics (in particular, generalizing the results obtained in \cite{vandeBruck:2025aaa}),
third,
in computing the effective potential for the bumblebee field, in close analogy with Ref.~\cite{ourbumb2}, thereby connecting the UV counterterm structure found here with the dynamical generation of LV vacua. We expect to perform these studies in forthcoming papers.

\acknowledgments
R. B. A. and W. C. were partially supported by the Coordena\c{c}\~ao de Aperfei\c{c}oamento de Pessoal de N\'ivel Superior (CAPES).  The work of A. Yu. P. was partially supported by the Conselho Nacional de Desenvolvimento Cient\'ifico e Tecnol\'ogico (CNPq),  Grant No.~303777/2023-0.  The work of A. C. L. was partially supported by CNPq, Grants No.~404310/2023-0 and No.~301256/2025-0.  P. J. P. would like to thank the Brazilian agency CNPq for financial support (PQ--2 grant, process No. 307628/2022-1).

\end{document}